\documentclass{kapproc}
\usepackage{t1enc} 
\usepackage{procps}
\usepackage[dvips]{graphicx}
\usepackage{amssymb} 
\usepackage{epsf}
\upperandlowercase
\kluwerbib

\newcommand{\hi}{H\,{\sc i}}

\newcommand{\mgii}{Mg\,{\sc ii}}

\newcommand{\mhi}{\mbox{$M_{\rm HI}$}}

\newcommand{\nhi}{\mbox{$N_{\rm HI}$}}
\newcommand{\fnhi}{\mbox{$f(N_{\rm HI})$}}

\newcommand{\ohi}{\mbox{$\Omega_{\rm HI}$}}

\newcommand{\icmsq}{\mbox{$ \rm cm^{-2}$}}

\begin{document}

\articletitle[Local galaxies as DLA analogs]{Local Galaxies as Damped Ly-$\alpha$ Analogs}

\author[Zwaan et al.] {M.~A.  Zwaan,\altaffilmark{1*}
J.~M. van der Hulst,\altaffilmark{2}
           F.~H. Briggs,\altaffilmark{3,4}
            M.~A.~W. Verheijen,\altaffilmark{2} 
 and E.~V. Ryan-Weber\altaffilmark{5} }

\altaffiltext{1}{ESO, Karl-Schwarzschild-Str. 2, 85748 Garching     b. M{\"u}nchen, Germany.}
\altaffiltext{2}{Kapteyn Astronomical Institute, PO box 800, 9700 AV Groningen, The Netherlands}
\altaffiltext{3}{RSAA, Mount Stromlo Observatory, Cotter     Road, Weston, ACT 2611, Australia}
\altaffiltext{4}{Australian National Telescope Facility, PO Box 76, Epping, NSW 1710, Australia}
\altaffiltext{5}{Institute of Astronomy, University of Cambridge, Madingley Road, Cambridge      CB30HA, UK}

\email{$^{*}$mzwaan@eso.org}

\begin{abstract}
We calculate in detail the expected properties of low redshift DLAs under the assumption that they arise in the gaseous disks of galaxies like those in the $z\approx 0$ population.  A sample of 355 nearby galaxies is analysed, for which high quality \hi\ 21-cm emission line maps are available as part of an extensive survey with the Westerbork telescope (WHISP).
We find that expected luminosities, impact parameters between quasars and DLA host galaxies, and metal abundances are in good agreement with the observed properties of 
DLAs and DLA galaxies. The measured redshift number density of $z=0$ gas above
the DLA limit is $dN/dz=0.045\pm 0.006$, which compared to higher $z$ measurements
implies that there is no evolution in the comoving density of DLAs along a line 
of sight between $z\sim 1.5$ and $z=0$, and a decrease of only a factor of two from $z\sim 4$ to the present time.
We conclude that the local galaxy population can explain all properties of low redshift DLAs.
\end{abstract}

\begin{keywords}
galaxies: ISM; (galaxies:) quasars: absorption lines; galaxies: evolution
\end{keywords}

\section{Introduction} The range of \hi\ column densities typically seen in routine 21-cm emission line observations of the neutral gas disks in nearby galaxies is very similar to those that characterise the Damped Lyman-$\alpha$ Systems or DLAs with $\nhi>2\times 10^{20}~\icmsq$. An attractive experiment would therefore be to map the \hi\ gas of DLA absorbing systems in 21-cm emission, and measure the DLAs' total gas mass, the extent of the gas disks and their dynamics. This would provide a direct observational link between DLAs and local galaxies, but unfortunately such studies are impossible with present technology (see e.g., Kanekar et al. 2001). The transition probability of the hyperfine splitting that causes the 21-cm line is extremely small, resulting in a weak line that can only be observed in emission in the very local ($z<0.2$) universe, with present technology. On the other hand, the identification of DLAs as absorbers in background QSO spectra is, to first order, not distance dependent because the detection efficiency depends mostly on the brightness of the background source, not on the redshift of the absorber itself. In fact, the lowest redshift ($z<1.7$) Lyman-$\alpha$ absorbers cannot be observed from the ground because the Earth's atmosphere is opaque to the UV wavelength range in which these are to be found. Furthermore, due to the expansion of the universe the redshift number density of DLAs decreases rapidly toward lower redshifts. Consequently, there are not many DLAs known whose 21-cm emission would be within the reach of present-day radio telescopes.

So, we are left with a wealth of information on the cold gas properties in local galaxies,
which has been collected over the last half century, and several hundreds DLA absorption profiles
at intermediate and high redshift,  but little possibility to bridge these two sets of information.
Obviously, most observers resort to the optical wavelengths to study DLAs but
attempts to directly image their host galaxies have  been notably unsuccessful (see e.g., Warren et al. 2001 and M{\o}ller et al. 2002 for reviews). A few positive identifications do exist, mostly the result of HST imaging.
%Possibly the best high redshift data to date are available for three objects imaged with STIS
%by M{\o}ller et al. (2002). They found emission from three DLA galaxies with
%spectroscopic confirmation and concluded that the objects are consistent with being
%drawn from the population of Lyman-break galaxies. 

Although the absolute number of DLAs at low $z$ is small, the success rate  for finding low-$z$ host galaxies is better for obvious reasons: the host galaxies are expected to be brighter and the separation on the sky between the bright QSO and the DLA galaxy is
likely larger. Early surveys for low-$z$ DLA host galaxies consisted of broad band imaging
and lacked spectroscopic follow-up (e.g., Le Brun et al.1997). Later studies aimed at measuring 
redshifts to determine the association of optically identified  galaxies with DLAs, either
spectroscopically (e.g., Rao et al. 2003), or using photometric redshifts (Chen \& Lanzetta 2003).
All together,  there are now $\sim 20$ DLA galaxies known at $z<1$. 
The galaxies span a wide range in galaxy properties, ranging from  inconspicuous LSB dwarfs to giant spirals and even early type galaxies. Obviously, it is not just the luminous, high surface brightness spiral galaxies that contribute to the \hi\ cross section above the DLA threshold. As explained above, we cannot study these galaxies in the 21-cm line on a case-by-case basis, but we can do a study of a statistical nature to see if the properties of
DLAs and DLA galaxies agree with our knowledge of \hi\ in the local universe.

\section{140,000 ``DLAs'' at $z=0$}
Blind 21-cm emission line surveys in the local universe with single dish radio telescopes such as Parkes or Arecibo have resulted in an accurate 
measurement of $\Omega_{\rm HI}(z=0)$, which can be used as a reference point 
for higher redshift DLA studies. $\Omega_{\rm HI}$ is simply calculated by integrating over the \hi\ mass function of galaxies, which is measured with surveys such as HIPASS
%The shape of the \hi\ mass function has not changed significantly since the early measurements from the late 1990, but the accuracy of the measurement has improved tremendously with HIPASS 
(Zwaan et al. 2005a). However, due to the large beam widths of the singe dish instruments, these surveys at best only barely resolve the detected galaxies and are therefore not very useful in constraining the column density 
distribution function of $z\approx 0$ \hi.
Hence, for this purpose we use the high resolution 21-cm maps of a large sample  of local galaxies
that have been observed with the Westerbork Synthesis
Radio Telescope. This sample is known as WHISP (van der Hulst et al. 2001) and consists of 355 galaxies
spanning a large range in \hi\ mass and optical luminosity. 
%In total, the equivalent of 
%more than one year of continuous observing has been devoted to obtain \hi\
%data cubes of these galaxies. 
The total number of independent column density measurements
above the DLA limit is $\sim 140,000$, which implies that the data volume of our present study is the equivalent of  $\sim 140,000$ DLAs at $z=0$!

Each galaxy in the sample is weighted according to the \hi\ mass function of galaxies.
%, which 
%expresses the space density of galaxies as a function of their \hi\ mass. 
We can now 
calculate the column density distribution function \fnhi, 
%which  
%is defined such that $f(\nhi) d\nhi dX$ is the number of absorbers with
%column density between $\nhi$ and $\nhi+d\nhi$ over an absorption distance interval $dX$. 
%At $z=0$ we can equate $dX$ to $dz$, such that \fnhi\ can be expressed as
\begin{equation}
f(\nhi) = \frac{c}{H_0} \frac{\int  \Phi(\mhi)  \Sigma (\nhi,\mhi) \,d\mhi}{d\nhi},
%f(\nhi) = \frac{c}{H_0} \frac{\int  \Phi({\bf x}_i)  \Sigma (\nhi,{\bf x}_i) \,d{\bf x}_i}{d\nhi},
\end{equation}
where $\Sigma(\nhi,\mhi)$ is the area function that describes for galaxies with \hi\ mass 
\mhi\ the area in $\rm Mpc^{-2}$ corresponding to a column density in the range \nhi\ to $\nhi+d\nhi$, and  $\Phi(\mhi)$ is the \hi\ mass function. $c/H_0$ converts the number of systems per Mpc to that per unit redshift.
%
%
%The integral sign denotes a summation over the whole range of \mhi\ of the galaxies in the sample. Finally, $c/H_0$ converts the number of systems per Mpc to that per unit redshift. 

\begin{figure}[t]
 \includegraphics[width=12.0cm,trim=2.0cm 11.05cm 1.8cm 1.8cm]{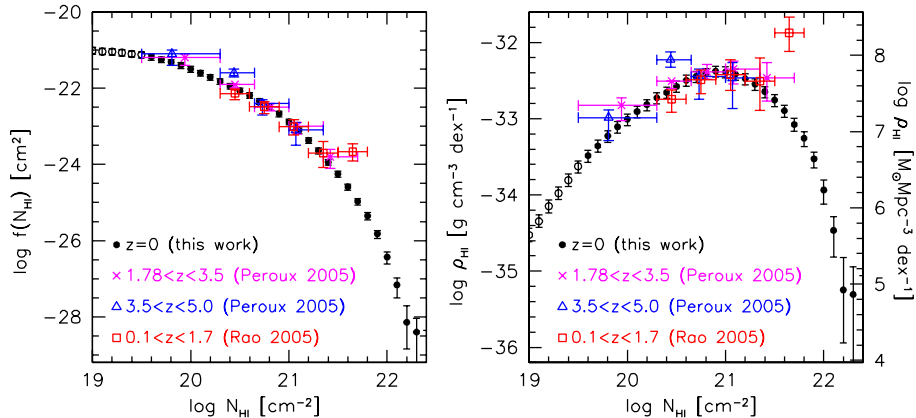}
  \caption{{\em Left:\/} The \hi\ column density distribution function \fnhi. The solid dots are from our analysis of 21-cm maps of local galaxies, the other points are from various surveys at higher redshifts as indicated in the legend. All calculations are based on a  $\Omega_{\rm m}= 0.3$, $\Omega_{\Lambda}=0.7$ cosmology. {\em Right:\/} The contribution of different column densities (per decade) to the integral \hi\ mass density. Note the very weak evolution in \ohi\ from high $z$ to the present.
  \label{whispfn2.fig}}
\end{figure}

Figure~\ref{whispfn2.fig} shows the resulting \fnhi\ on the left, and the derived \hi\ mass density per decade of \nhi\ on the right. For comparison with higher redshift observations, we also plot the results from two other studies.
% P{\'e}roux et al. (2005), Rao et al. (2005), and Prochaska et al. (2005).  
The P{\'e}roux (2005) measurements of \fnhi\ below the DLA limit are the result of their new UVES survey for ``sub-DLAs''. 
The intermediate redshift points from Rao et al. (2005) are based on \mgii-selected DLA
systems. The surprising result from this figure is that there appears to be only very mild evolution in the intersection cross section of \hi\ from redshift $z\sim 5$ to the present. 
From this figure we can determine the redshift number density of $\log\nhi>20.3$ gas and 
find that $dN/dz=0.045\pm0.006$, in good agreement with earlier measurements at $z=0$. Compared to the most recent measurements of $dN/dz$ at intermediate and high $z$, this implies that the comoving
number density (or the ``space density times cross section'') of DLAs does not evolve after $z\sim 1.5$. In other words, the local galaxy population explains the incidence rate of 
low and intermediate $z$ DLAs and there is no need for a population of hidden very low surface brightness (LSB) galaxies or isolated \hi\ clouds (dark galaxies).

The right hand panel shows that at $z=0$ most of the \hi\ atoms are in column densities
around $\nhi=10^{21}~\icmsq$. This also seems to be the case at higher redshifts, although
the distribution might flatten somewhat. 
The one point that
clearly deviates is the highest  \nhi\ point from Rao et al. (2005) at $\log\nhi=21.65$. 
%This elevated interception rate of high \nhi\ \mgii-selected intermediate redshift DLAs
%is also reported in their earlier work. 
The figure very
clearly demonstrates that this point dominates the \ohi\ measurement at intermediate
redshifts. 
It is therefore important to understand whether the \mgii-based results really indicate
that high column densities ($\log \nhi\sim21.65$) are rare at high and low redshift, but much more ubiquitous at intermediate redshifts, or whether the \mgii\ selection introduces currently unidentified biases.

\section{Expected properties of low-$z$ DLAs}

\begin{figure}[t]
 \includegraphics[width=6.0cm,trim=0.cm 7.9cm 5.1cm 0.5cm]{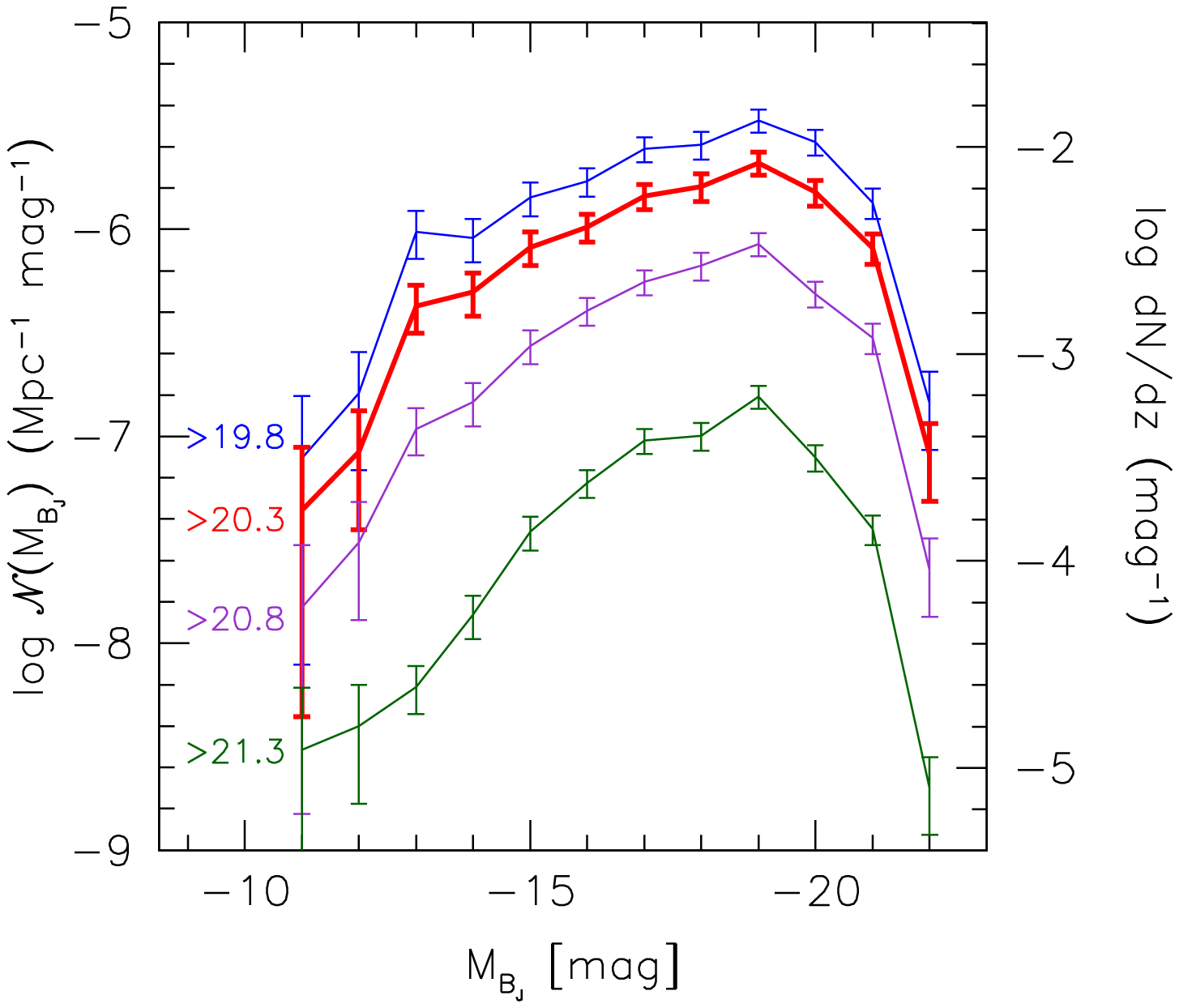}
 \includegraphics[width=6.0cm,trim=0.cm 7.9cm 5.1cm 0.5cm]{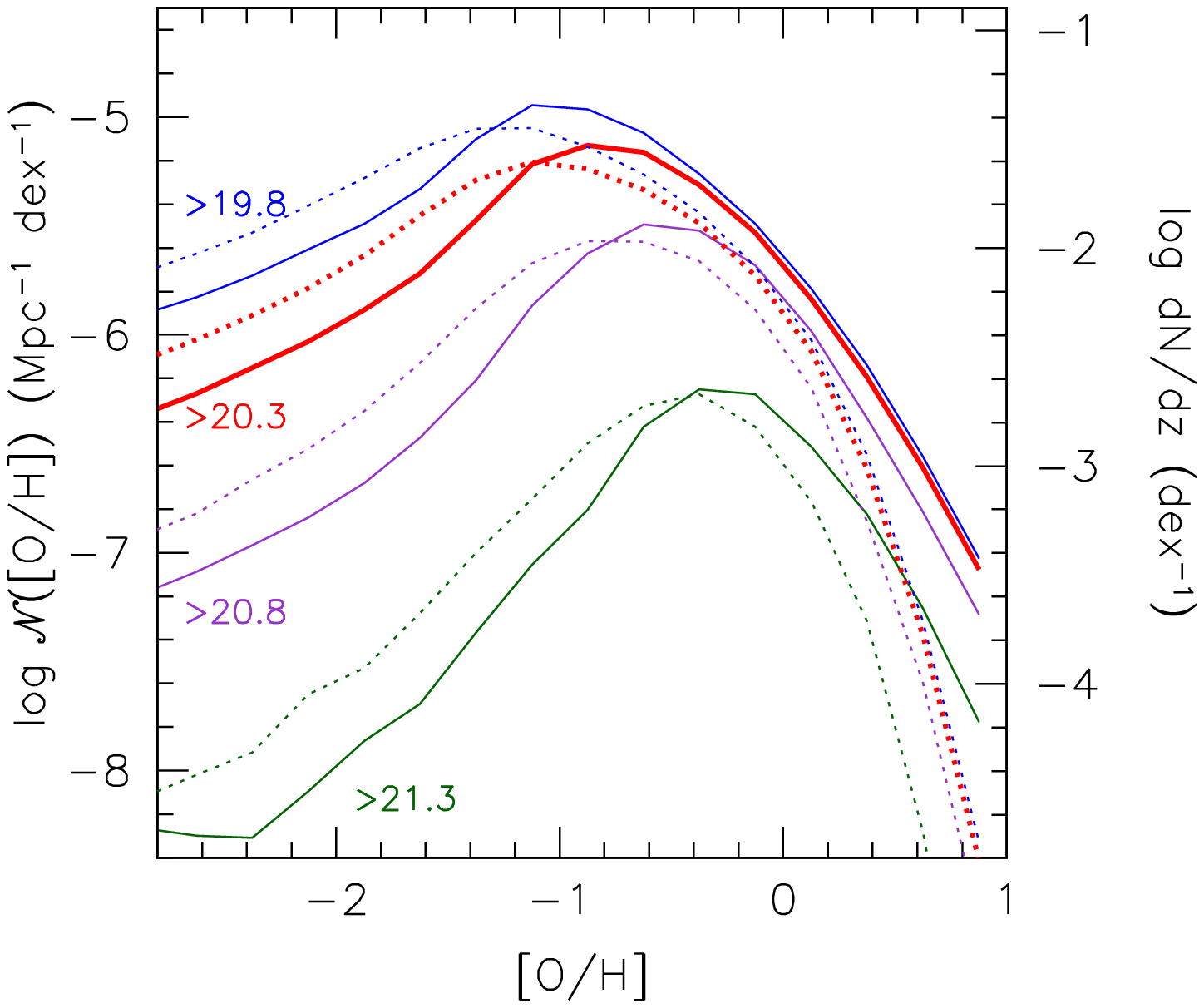}
  \caption{{\em Left:\/} The expected distribution of $B$-band absolute magnitudes 
  of $z=0$ high \hi\ column density systems. The lines plus errorbars show the product of cross sectional area and space density, which translates to the number of expected absorbers per Mpc per magnitude. The right axis shows the corresponding number of absorbers per unit redshift $dN/dz$. The different lines correspond to different column density limits, as indicated by the labels. The thick line corresponds to the classical DLA limit of $\log \nhi>20.3$.
  {\em Right:\/} The probability distribution of oxygen abundance [O/H] of \hi\ absorbers. 
The solid lines refer to the approach of assuming fixed [O/H] gradients and the the dashed lines refer to varying gradients (see text).    
    \label{pdfs.fig}}
\end{figure}

Now that we have accurate cross section measurement of all galaxies in our sample, and know what the space density of our galaxies is, we can calculate the cross-section weighted probability distribution functions of various galaxy parameters. %In other words, we can calculate the probability distribution of parameters of galaxies whose gas disks are encountered along random lines of sight  in the local universe.
Figure~\ref{pdfs.fig} shows two examples. The left panel shows the $B$-band absolute magnitude distribution of cross-section selected galaxies above four different \hi\ column density cut-offs. 
87\% of the DLA cross-section appears to be in galaxies that are fainter than $L_*$, and 45\% is in galaxies with $L<L_*/10$. 
These numbers agree very well with the luminosity distribution of
$z<1$ DLA host galaxies. Taking into account the non-detections of DLA
host galaxies and assuming that these are $\ll L_*$, we find that 80\%
of the $z<1$ DLA galaxies are sub-$L_*$.
The median absolute magnitude of a $z=0$ DLA galaxy is expected to be $M_B=-18.1$ ($\sim L_*/7$). The conclusion to draw from this is that we should not be surprised to find that identifying  DLA host galaxies is difficult. Most of them (some 87\%) are expected to be sub-$L_*$
and many are dwarfs.

Using similar techniques, we find that the expected median impact parameter of $\log \nhi>20.3$ systems is 7.8 kpc, 
whereas the  median impact parameter of identified $z<1$ DLA galaxies is 8.3 kpc.
Assuming no evolution in the properties of galaxies' gas disks, these numbers imply that 37\% of the impact parameters are expected to be less than $1''$ for systems at $z=0.5$. 
This illustrates that very high spatial resolution imaging programs
are required to successfully identify a typical DLA galaxy at intermediate redshifts.

The right panel in Figure~\ref{pdfs.fig} shows the probability distribution of oxygen abundance in $z=0$ DLAs. We constructed this diagram by assigning to every \hi\ pixel in our 21-cm maps an oxygen abundance, based on the assumption that the galaxies in our sample follow the local metallicity--luminosity ($Z-L$) relation (e.g. Garnett 2002), and that each disk
shows an abundance gradient of [O/H] of $-0.09 \rm \,dex\, kpc^{-1}$ (e.g., Ferguson et al. 1998) along the major axis.
The solid lines correspond to these assumptions, the dotted lines are 
for  varying metallicity gradients in disks of different absolute brightness. 
The main conclusion is that the metallicity distribution for \hi\ column densities $\log \nhi>20.3$
peaks around [O/H]=$-1$ to $-0.7$, much lower than the mean value of an $L_*$ galaxy of
[O/H]$\approx 0$. The reason for this being that {\em 1)\/} much of the DLA cross section is in sub-$L_*$ galaxies, which mostly have sub-solar metallicities, and  {\em 2)\/} for the more luminous, larger galaxies, the highest interception probability is at larger impact parameters from the centre, where the metallicity is lower.
%Furthermore, we see a strong correlation between the peak of the [O/H] probability distribution and the \hi\ column density limit: for \hi\ column densities in excess of $\log \nhi=21.3$, the expected peak of
%the distribution is at [O/H]=$-0.2$. This correlation is expected, since our assumed radial abundance gradients impose a relation between \nhi\ and [O/H]. 
%Taking into account uncertainties in the \hi\ mass function, in the $L-Z$ relation and in the abundance gradients in galaxies, we adopt a value of [O/H]$=-0.85\pm 0.2$ (1/7 solar) as a representative value for the median cross-section--weighted abundance of \hi\ gas in the local universe above the DLA column density limit.
%For the mean mass-weighted metallicity of \hi\ gas with $\log \nhi>20.3$ at $z=0$ we find the value of  $\log {Z}=-0.35\pm 0.2$. 
Interestingly, this number is very close to the $z=0$ extrapolations of 
metallicity measurements in DLAs at higher $z$ from Prochaska et al. (2003) and Kulkarni et al. (2005).  For the mean mass-weighted metallicity of \hi\ gas with $\log \nhi>20.3$ at $z=0$ we find the value of  $\log {Z}=-0.35\pm 0.2$, also consistent with 
the $z=0$ extrapolation of \nhi-weighted metallicities in DLAs,
although we note this extrapolation has large uncertainties given the
poor statistics from DLAs at $z<1.5$.
%These numbers are not very different from the $z=0$ extrapolations of metallicity measurements in DLAs. Prochaska et al. (2003) recently compiled metallicity measurements in DLAs over a redshift range $0<z<4.5$, mostly based on $\alpha$-elements. The $z=0$
%extrapolations of the least squares fits to these data, give an \nhi-weighted value of 
%$\log Z\approx -0.5$ and median value of $\log Z\approx  -0.7$. These numbers are within %$1\sigma$ of our measurements of $-0.35 \pm 0.1$ and $-0.85\pm 0.2$, respectively.
These results 
are in  good agreement with the hypothesis 
that DLAs arise in the \hi\ disks of galaxies.
%provided that the metallicities in DLAs continue to evolve since $z=0.5$ with a rate similar to that observed between $z=0.5$ and $z=4$.

\section{Conclusions}

The local galaxy population can explain the incidence rate and metallicities of DLAs, the luminosities of their host galaxies, and the impact parameters between centres of host galaxies and the background QSOs.

This work is presented in much more detail in a forthcoming paper (Zwaan et al. 2005b).

%\begin{acknowledgments}

%\end{acknowledgments}

%\begin{thebibliography}{}

%\end{thebibliography}


\begin{chapthebibliography}{1}
%\bibitem[\protect\citeauthoryear{Chen, Kennicutt, \& 
%Rauch}{2005}]{chen2005} Chen H., Kennicutt Jr R.~C., Rauch M., 
%2005, astro-ph/0411006 
\bibitem[Chen \& Lanzetta(2003)]{chen2003} Chen, H.~\& Lanzetta, 
K.~M.\ 2003, ApJ, 597, 706 
%\bibitem[\protect\citeauthoryear{Driver et al.}{2004}]{driver2004} 
%Driver, S., et al., 2004, MNRAS, in press
\bibitem[\protect\citeauthoryear{Ferguson, Gallagher, \& 
Wyse}{1998}]{ferguson1998} Ferguson A.~M.~N., Gallagher J.~S., Wyse 
R.~F.~G., 1998, AJ, 116, 673 
\bibitem[\protect\citeauthoryear{Garnett}{2002}]{garnett2002} 
Garnett D.~R., 2002, ApJ, 581, 1019 
\bibitem[\protect\citeauthoryear{Kanekar et 
al.}{2001}]{kanekar2001} Kanekar N., Chengalur J.~N., Subrahmanyan 
R., Petitjean P., 2001, A\&A, 367, 46 
\bibitem[\protect\citeauthoryear{Kulkarni et 
al.}{2005}]{kulkarni2005} Kulkarni V.~P., Fall S.~M., Lauroesch 
J.~T., York D.~G., Welty D.~E., Khare P., Truran J.~W., 2005, ApJ, 618, 68 
\bibitem[Le Brun et al.(1997)]{lebrun1997} Le Brun, V., Bergeron, J., Boisse, 
P., \& Deharveng, J.~M.\ 1997, A\&A, 321, 733 
\bibitem[M{\o}ller et al.(2002)]{moller2002} M{\o}ller, P., 
Warren, S.~J., Fall, S.~M., Fynbo, J.~U., \& Jakobsen, P.\ 2002, ApJ, 574, 
51 
\bibitem[\protect\citeauthoryear{M{\o}ller, Fynbo, \& 
Fall}{2004}]{moller2004} M{\o}ller P., Fynbo J.~P.~U., Fall 
S.~M., 2004, A\&A, 422, L33 
%\bibitem[P{\' e}roux et al.(2003)]{peroux2003} P{\' e}roux, C., McMahon, R.~G., 
%Storrie-Lombardi, L.~J., \& Irwin, M.~J.\ 2003, MNRAS, 346, 1103 
\bibitem[P{\' e}roux et al.(2005)]{peroux2005} P{\' e}roux, C., 
Dessauges-Zavadsky, M., D'Odorico, S., Kim, T., \& McMahon, R.~G.\ 2005, MNRAS, 
{\em in press}
\bibitem[\protect\citeauthoryear{Prochaska et 
al.}{2003}]{prochaska2003} Prochaska J.~X., Gawiser E., Wolfe A.~M., 
Castro S., Djorgovski S.~G., 2003, ApJ, 595, L9 
%\bibitem[Prochaska et al.(2005)]{prochaska2005} Prochaska, 
%J.~X.~, Herbert-Fort, S. \& Wolfe, A.~M.\ 2005, ApJ, {\em submitted}
%\bibitem[Rao \& Turnshek(2000)]{rao2000} Rao, S.~M.~\& 
%Turnshek, D.~A.\ 2000, ApJS, 130, 1 
\bibitem[Rao et al.(2003)]{rao2003} Rao, S.~M., Nestor, D.~B., 
Turnshek, D.~A., Lane, W.~M., Monier, E.~M., \& Bergeron, J.\ 2003, ApJ, 
595, 94 
%\bibitem[Rao et al.(2005)]{rao2005} Rao, S.~M.,
%Turnshek, D.~A., \& Nestor, D.~B.\ 2005, ApJ, {\em submitted}
\bibitem[]{}Rao, S.~M. 2005, astro-ph/0505479
\bibitem[\protect\citeauthoryear{van der Hulst, van Albada, \& 
Sancisi}{2001}]{vanderhulst2001} van der Hulst J.~M., van Albada T.~S., 
Sancisi R., 2001, ASP Conf.~Ser.~240: Gas and 
Galaxy Evolution, 240, 451 
%\bibitem[\protect\citeauthoryear{Vila-Costas \& 
%Edmunds}{1992}]{vila-costas1992} Vila-Costas M.~B., Edmunds M.~G., 
%1992, MNRAS, 259, 121 
\bibitem[\protect\citeauthoryear{Warren et al.}{2001}]{warren2001} 
Warren S.~J., M{\o}ller P., Fall S.~M., Jakobsen P., 2001, MNRAS, 326, 759 
%\bibitem[Zwaan et al.(2003)]{zwaan2003} Zwaan, M.~A., et al.\ 
%2003, AJ, 125, 2842 
\bibitem[Zwaan et al.(2005)]{zwaan2005} Zwaan, M.~A., et al.\ 
2005a, MNRAS, 359, L30
\bibitem[Zwaan et al.(2005)]{zwaan2005b} Zwaan, M.~A., van der Hulst, J.~M., Briggs, F.~H., Verheijen, M.~A.~W., Ryan-Weber, E.~V., 2005b, MNRAS, {\em submitted}

\end{chapthebibliography}
\end{document}